\documentclass[12pt,letterpaper]{article}

\usepackage{epsfig}
\usepackage{multirow}
\usepackage[hypertex]{hyperref} 

\newcommand{\bdm}{\begin{displaymath}}
\newcommand{\edm}{\end{displaymath}}
\newcommand{\beq}{\begin{equation}}
\newcommand{\eeq}{\end{equation}}
\newcommand{\bea}{\begin{eqnarray}}
\newcommand{\eea}{\end{eqnarray}}

\newcommand{\Mpi}{M_\pi}
\newcommand{\Fpi}{F_\pi}
\newcommand{\Mka}{M_K}
\newcommand{\Fka}{F_K}

\newcommand{\mr}{\mathrm}
\newcommand{\ovr}{\over}
\newcommand{\MeV}{\,\mr{MeV}}
\newcommand{\fm}{\,\mr{fm}}

\def\reff#1{\ref{#1}}
\def\eq#1{Eq.\,(\reff{#1})}
\def\fig#1{Fig.\,\reff{#1}}
\def\sec#1{Sec.\,\reff{#1}}
\def\tab#1{Tab.\,\reff{#1}}

\def\lsim{\raise0.3ex\hbox{$<$\kern-0.75em\raise-1.1ex\hbox{$\sim$}}}
\def\gsim{\raise0.3ex\hbox{$>$\kern-0.75em\raise-1.1ex\hbox{$\sim$}}}

\begin{document}


\begin{center}
{\LARGE\bf The ratio $F_K/F_\pi$ in QCD}
\end{center}
\vspace{10pt}

\begin{center}
  S.\,D\"urr$^{1,2}$, Z.\,Fodor$^{1,2,3}$, 
  C.\,Hoelbling$^1$, S.\,D.\,Katz$^{1,3}$, S.\,Krieg$^5$,
  T.\,Kurth$^1$, L.\,Lellouch$^4$, T.\,Lippert$^{1,2}$, A.\,Ramos$^4$,
  K.\,K.\,Szab\'o$^1$
  \\[4mm]
  [Budapest-Marseille-Wuppertal Collaboration]
  \\[4mm]
  $^1${\sl Bergische Universit\"at Wuppertal, Gaussstr.\,20, D-42119
    Wuppertal,
    Germany}\\
  $^2${\sl J\"ulich Supercomputing Center, Forschungszentrum J\"ulich,
    D-52425
    J\"ulich, Germany}\\
  $^3${\sl Institute for Theoretical Physics, E\"otv\"os University,
    H-1117
    Budapest, Hungary}\\
  $^4${\sl Centre de Physique Th\'eorique%
    \footnote[0]{Centre de Physique Th\'eorique is ``UMR
      6207 du CNRS et des universit\'es d'Aix-Marseille I,
      d'Aix-Marseille II et du Sud Toulon-Var, affili\'ee \`a la
      FRUMAM''.}, CNRS Luminy, Case 907, F-13288 Marseille Cedex 9,
    France} $^5${\sl Center for Theoretical Physics, MIT, Cambridge,
    MA 02139-4307, USA}
\end{center}
\vspace{10pt}

\begin{abstract}
\noindent
We determine the ratio $F_K/F_\pi$ in QCD with $N_f=2+1$ flavors of
sea quarks, based on a series of lattice calculations with three
different lattice spacings, large volumes and a simulated pion mass
reaching down to about 190\,MeV. We obtain
$F_K/F_\pi=1.192(7)_\mr{stat}(6)_\mr{syst}$. This result is then used to
give an updated value of the CKM matrix element $|V_\mathrm{us}|$. The
unitarity relation for the first row of this matrix is found to be
well observed.
\end{abstract}
\vspace{10pt}

\clearpage


\section{Introduction}


An accurate determination of the Cabibbo-Kobayashi-Maskawa (CKM) matrix element
$|V_\mr{us}|$ is important, because it allows to put bounds on possible
extensions of the Standard Model of Particle Physics which, in turn, are
relevant to guide direct searches, e.g.\ those planned at the Large Hadron
Collider (LHC) at CERN \cite{LHC}.

Since the kaon is the lightest particle with strangeness, it is not surprising
that much of the recent progress in this field derives from precision studies
of leptonic and semi-leptonic decays of kaons (see e.g.\
\cite{Antonelli:2008jg}).
The theoretical challenge is to link the experimentally observed branching
fractions to fundamental parameters in the underlying theory, e.g.\ the CKM
matrix elements in the Standard Model.
In this step lattice QCD calculations can help by providing decay constants and
transition form factors.
For an overview on these activities see e.g.\ the summary talks on kaon physics
at the last three lattice conferences
\cite{Juttner:2007sn,Lellouch:2009fg,Lubicz:LAT09}.

In this paper we shall follow a proposal by Marciano \cite{Marciano:2004uf} to
derive $|V_\mr{us}|$ from $|V_\mr{ud}|$, using a lattice determination of the
ratio $F_K/F_\pi$ of leptonic decay constants.
More specifically, in
\beq
{\Gamma(K\to l\bar\nu_l)\over\Gamma(\pi\to l\bar\nu_l)}
=
{|V_\mr{us}|^2\over|V_\mr{ud}|^2}
{F_K^2\over F_\pi^2}
{M_K(1-m_l^2/M_K^2)^2\over M_\pi(1-m_l^2/M_\pi^2)^2}
\big\{
1+{\alpha\over\pi}(C_K-C_\pi)
\big\}
\label{marciano}
\eeq the l.h.s., even after dividing it by the radiative correction
factor (the last bracket on the r.h.s.), is known to 0.4\% precision,
if $l=\mu$ is considered \cite{Amsler:2008zz}.  Also $M_K$, $M_\pi$
and $m_\mu$ are known with a relative precision of
$3\!\cdot\!10^{-5}$, $3\!\cdot\!10^{-6}$ and $10^{-7}$, respectively
\cite{Amsler:2008zz}.  And $|V_\mr{ud}|$ has been determined from
super-allowed nuclear $\beta$-decays with an accuracy better than
$0.03\%$ \cite{Amsler:2008zz,Hardy:2008gy}.  Therefore, the limiting
factor for a precise determination of $|V_\mr{us}|$ via
(\ref{marciano}) is $F_K/F_\pi$ -- this ratio is typically determined
with a precision of a few percent in present days lattice QCD studies.

Here we present a state-of-the art determination of $F_K/F_\pi$ in QCD
in which all sources of systematic uncertainty are properly taken into
account.  We have performed a series of dynamical lattice calculations
with a degenerate up and down quark mass $m_{ud}$ and a separate
strange quark mass $m_s$, a scheme commonly referred to as $N_f=2+1$.
The strange quark mass is roughly held fixed close to its physical
value ($m_s\!\simeq\!m_s^\mr{phys}$), while the light quark mass is
heavier than in the real world, but varied all the way down to values
which make a controlled extrapolation to the physical mass possible.
The spatial size $L$ is chosen sufficiently large, such that the
extracted $F_K/F_\pi$ values can be corrected, by means of Chiral
Perturbation Theory, for (small) finite-volume effects.  Since our
calculation includes three lattice spacings, a combined fit to all
simulation results yields a controlled extrapolation to the continuum
and to the physical mass point.  Details of our action and algorithm
have been specified in \cite{Durr:2008rw}, where evidence for
excellent scaling properties is also given.

This paper is organized as follows.  Sec.\,2 contains a survey of
where our simulation points are located in the
$(M_\pi^2,2M_K^2-M_\pi^2)$ plane, a discussion of which chiral
extrapolation formulae might be appropriate, an account of our
strategies to quantify (and correct for) cutoff and finite-volume
effects, and a description of the overall fitting procedure that is
used to determine $F_K/F_\pi$.  In Sec.\,3 the final result is given
and compared to other unquenched calculations
\cite{Aoki:2002uc}-\nocite{Aubin:2004fs,Bazavov:2009bb,Beane:2006kx,
Follana:2007uv,Blossier:2009bx,
Allton:2008pn,Aoki:2008sm}\cite{Aubin:2008ie} of $F_K/F_\pi$.  In
Sec.\,4 our result is converted, by means of (\ref{marciano}), into a
value for $|V_\mr{us}|$.  We find that, within errors, the first row
unitarity relation $|V_\mr{ud}|^2+|V_\mr{us}|^2+|V_\mr{ub}|^2=1$ is
well observed.  This, in turn, puts tight constraints on possible
extensions of the Standard Model~\cite{LHC,Antonelli:2008jg}.


\section{Simulation and analysis details}


\subsection{Simulation parameters}

Our gauge and fermion actions, as well as details of the algorithm
that is used to simulate QCD with $N_f=2+1$ dynamical flavors, have
been specified in \cite{Durr:2008rw}.  Here, it is sufficient to say
that our action combines good scaling properties with a one-to-one
matching of lattice-to-continuum flavor; this avoids the complications
of a low-energy theory with unphysical flavor symmetry breaking.

\begin{figure}
\epsfig{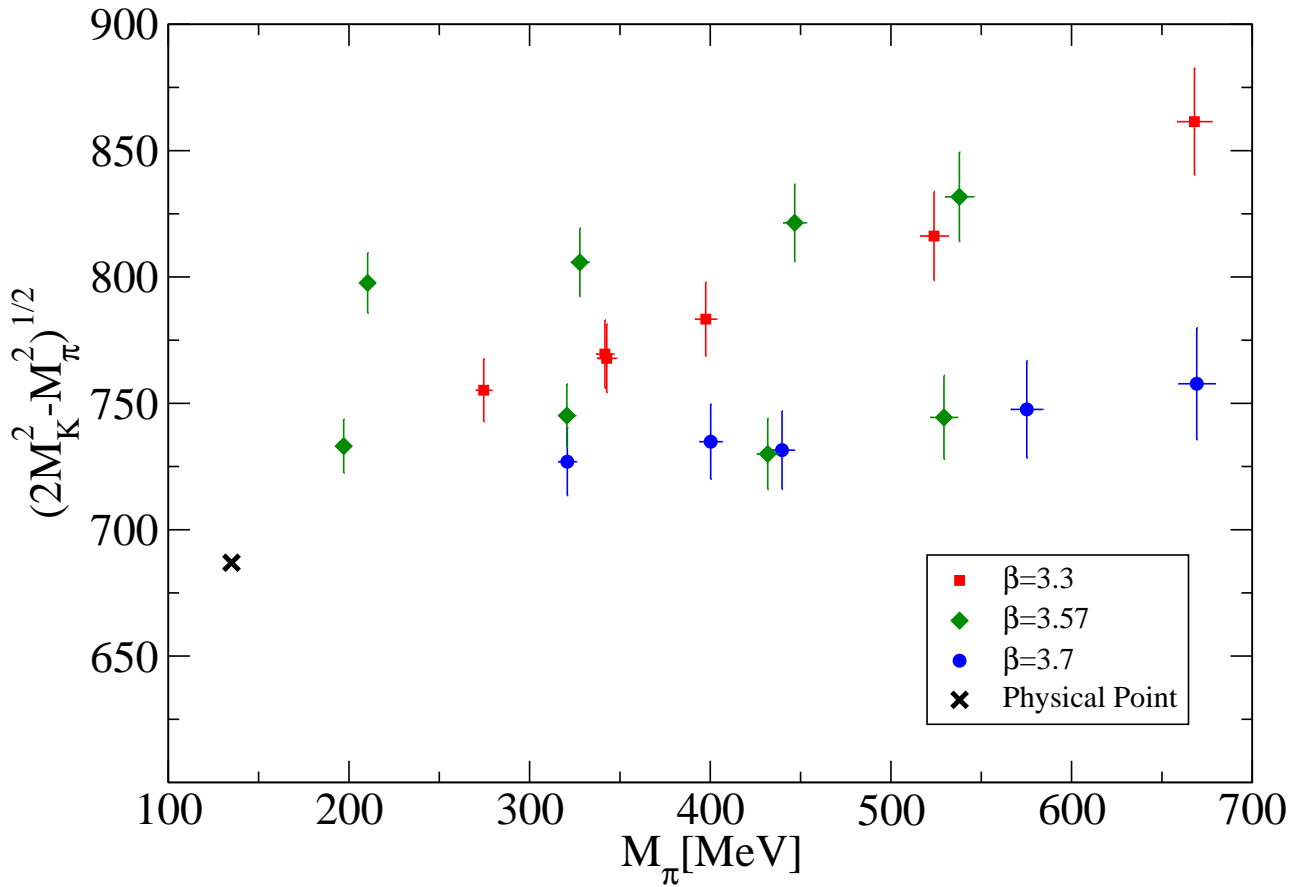}
\vspace{-8mm}
\caption{\sl Overview of our simulation points in terms of $\Mpi$
  and $\sqrt{2\Mka^2-\Mpi^2}$. The former gives a measure of the isospin
  averaged up and down quark mass while the later determines the
  strange quark mass. The symbols refer to the three lattice spacings
  $a\!\simeq\!0.124\fm$ $(\beta=3.3)$, $a\!\simeq\!0.083\fm$
  $(\beta=3.57)$, and $a\!\simeq\!0.065\fm$ $(\beta=3.7)$,
  respectively, and the physical point is marked with a
  cross. Error bars are statistical only. The numbers given
  correspond to one of the 18 two-point function, time fit-intervals
  that we use in our estimate of systematic uncertainties
  (see \sec{sec:fitting} for details). At $\beta=3.3$ and $\beta=3.7$
  only a single value of the strange quark mass, close to the physical
  value, is considered.  At $\beta=3.57$, simulations are performed at
  three values of the strange quark mass (see text for details).}
\label{fig1}
\end{figure}

We adjust the quark masses $m_{ud}$, $m_s$ and set the lattice spacing
$a$ using $aM_\pi$, $aM_K$ and $aM_\Xi$.  We adopt a mass independent
scale setting scheme.  This means that we extrapolate, for each
coupling $\beta=3.3,3.57,3.7$, the values $aM_\pi,aM_K,aM_\Xi$ to the
point where any two of the ratios in $M_\pi/M_K/M_\Xi$ agree with
their experimental values.  Alternatively we have used $\Omega$ instead
of $\Xi$ to set the scale and used the difference between the two
methods as an estimate of the systematics of scale fixing (see
\sec{sec:fitting}). Our simulations do not account for isospin
breaking and electromagnetic interactions; thus experimental inputs
need to be corrected for these effects.  We use
$\Mpi^\mr{phys}=135\MeV$, $\Mka^\mr{phys}=495\MeV$,
$M_\Xi^\mr{phys}=1318\MeV$ and $M_\Omega^\mr{phys}=1672\MeV$ with an
error of a few MeV~\cite{Aubin:2004fs}. 

Using LO chiral perturbation theory as a guide, we fix the bare
strange quark mass at a given $\beta$, such that the computed
$2\Mka^2-\Mpi^2$ approaches its physical value when extrapolated in
$M_\pi^2$ to the physical point. As \tab{tab1} indicates and
\fig{fig1} illustrates, only one bare strange quark mass is used at
$\beta\!=\!3.3$ and at $\beta\!=\!3.7$, and it is very near the value
required to reach the physical point.  At $\beta\!=\!3.57$, three
values of the strange quark mass are considered to provide some lever
arm for an interpolation to the physical $m_s$. The fact that our
simulation points lie above the physical value of $2\Mka^2-\Mpi^2$
is due in part to cutoff effects in the relation of this
quantity to the vector Ward identity strange quark mass. Indeed, for
$\beta\!=\!3.3$ and $3.7$, $2\Mka^2-\Mpi^2$ extrapolates near its
physical value at $\Mpi=\Mpi^\mr{phys}$, and the slope with which it
extrapolates tends to zero as $\beta$ increases.

Regarding $m_{ud}$, we cover pion masses all the way down to
$\sim\!190\MeV$.  On every ensemble the ratio $F_K/F_\pi$ is measured
with the valence up/down or strange quark mass set equal to the
corresponding sea quark mass, so only the unitary theory is
considered. Results are collected in \tab{tab1}.
It is worth noting that this same dataset was successfully used to
determine the light hadron spectrum~\cite{Durr:2008zz}.

\begin{table}
\centering\small
\begin{tabular}{@{\,}l@{\,}@{\,}c@{\,}@{\,}c@{\,}@{\,}c@{\,}@{\,}c@{\,}@{\,}c@{\,}@{\,}c@{\,}@{\,}c@{\,\,}}
\hline
\hline
$\;\beta$ & $am_{ud}$ & $am_s$ & $L^3\!\times\!T$ &
traj. & $aM_\pi$ & $aM_K$ & $F_K/F_\pi$\\
\hline
\multirow{6}{*}{3.3}
 & -0.0960 & -0.057 & $16^3\!\times\!32$ & 10000 & 0.4115(6) & 0.4749(6) & 1.0474(5) \\
 & -0.1100 & -0.057 & $16^3\!\times\!32$ &  1450 & 0.322(1) & 0.422(1) & 1.079(2) \\
 & -0.1200 & -0.057 & $16^3\!\times\!64$ &  4500 & 0.2448(9) & 0.3826(6) & 1.113(3) \\
 & -0.1233 & -0.057 & $24^3\!\times\!64$ &  2000 & 0.2105(8) & 0.3668(6) & 1.137(6) \\
 & -0.1233 & -0.057 & $32^3\!\times\!64$ &  1300 & 0.211(1) & 0.3663(8) & 1.130(5)\\
 & -0.1265 & -0.057 & $24^3\!\times\!64$ &  2100 & 0.169(1) & 0.3500(7) & 1.153(6) \\
\hline
\multirow{4}{*}{3.57}
 & -0.0318 & 0,-0.010 & $24^3\!\times\!64$ & 1650,1650 & 0.2214(7),0.2178(5) & 0.2883(7),0.2657(5) & 1.085(1),1.057(1) \\
 & -0.0380 & 0,-0.010 & $24^3\!\times\!64$ & 1350,1550 & 0.1837(7),0.1778(7) & 0.2720(6),0.2469(6) & 1.107(3),1.092(2) \\
 & -0.0440 & 0,-0.007 & $32^3\!\times\!64$ & 1000,1000 & 0.1348(7),0.1320(7) & 0.2531(6),0.2362(7) & 1.154(4),1.132(4) \\
 & -0.0483 & 0,-0.007 & $48^3\!\times\!64$ &  500,1000 & 0.0865(8),0.0811(5) & 0.2401(8),0.2210(5) & 1.22(1),1.178(7) \\
\hline
\multirow{5}{*}{3.7}
 & -0.007 & 0.0 & $32^3\!\times\!96$ & 1100 & 0.2130(4)  & 0.2275(4)  & 1.0223(4) \\
 & -0.013 & 0.0 & $32^3\!\times\!96$ & 1450 & 0.1830(4)  & 0.2123(3)  & 1.0476(9) \\
 & -0.020 & 0.0 & $32^3\!\times\!96$ & 2050 & 0.1399(3)  & 0.1920(3)  & 1.089(2) \\
 & -0.022 & 0.0 & $32^3\!\times\!96$ & 1350 & 0.1273(5)  & 0.1882(4)  & 1.102(2) \\
 & -0.025 & 0.0 & $40^3\!\times\!96$ & 1450 & 0.1021(4)  & 0.1788(4)  & 1.137(6) \\
\hline
\hline
\end{tabular}
\caption{\sl Parameters and selected results from our simulations.
  The errors quoted here are purely statistical. These results
  correspond to one of the 18 two-point function, time fit-intervals
  that we use in our estimate of systematic uncertainties
  (see \sec{sec:fitting} for details). In this
  particular analysis, the scales at $\beta=3.3, 3.57, 3.7$ are
  $a^{-1}=1616(20)\MeV$, $2425(27)\MeV$, $3142(37)\MeV$, respectively.}
\label{tab1}
\end{table}

We now give details of how we extrapolate to the physical mass point, to the
continuum, and to infinite volume. We continue with an explanation of our
global fitting strategy and the pertinent assessment of systematic errors.


\subsection{Extrapolation to the physical mass point}
\label{sec:massextrap}

From the discussion above it is clear that we need to extrapolate our
results for $F_K/F_\pi$ in $m_{ud}$ to $m_{ud}^\mr{phys}$, while the
strange quark mass is already close to $m_s^\mr{phys}$. It is
important to note that even the extrapolation in $m_{ud}$ is small,
amounting to only a couple percent in $F_K/F_\pi$. Thus, any
reasonable, well-motivated functional form which fits our results with
a good confidence level should give a reliable estimate of
$F_K/F_\pi$ at the physical point.  In order to assess the theoretical
error that arises in this extrapolation, we choose to invoke three
frameworks to parameterize the quark mass dependence:
\begin{itemize}
\itemsep -2pt
\vspace{-4pt}
\item[({\it i}\,)]
$SU(3)$ chiral perturbation theory ($\chi$PT)~\cite{Gasser:1984gg},
\item[({\it ii}\,)]
heavy kaon $SU(2)$ chiral perturbation theory~\cite{Allton:2008pn},
\item[({\it iii}\,)]
``Taylor'' extrapolations involving polynomial ansaetze~\cite{Lellouch:2009fg}.
\vspace{-4pt}
\end{itemize}
We now summarize the theoretical background of these functional forms.

Ad ({\it i}\,): $SU(3)$ $\chi$PT assumes that the $u$, $d$ and $s$
quark masses are small compared to the scale of chiral symmetry
breaking. At NLO, the ratio $F_K/F_\pi$ can be written
\cite{Gasser:1984gg}
\beq
{\Fka\ovr\Fpi} =
{1-{1\ovr32\pi^2F_0^2}
\Big\{{3\ovr4}\Mpi^2\log({\Mpi^2\ovr\mu^2})+
{3\ovr2}\Mka^2\log({\Mka^2\ovr\mu^2})+[\Mka^2\!-\!{1\ovr4}\Mpi^2]
\log({4\Mka^2-\Mpi^2\ovr3\mu^2})\Big\}
+{4\ovr F_0^2}\Mka^2L_5
\ovr
1-{1\ovr32\pi^2F_0^2}
\Big\{2\Mpi^2\log({\Mpi^2\ovr\mu^2})+\Mka^2\log({\Mka^2\ovr\mu^2})\Big\}
+{4\ovr F_0^2}\Mpi^2L_5}
\label{ratio2}
\eeq
where, for $P=\pi,K,\eta$,
\beq
\mu_P={M_P^2\ovr32\pi^2F_0^2}\log({M_P^2\over\mu^2})
\eeq
and where $M_P$, at this order, can be taken to be the lattice
measured masses, with $M_\eta^2\stackrel{\mr{LO}}{=}(4
M_K^2-M_\pi^2)/3$, using the LO $SU(3)$ relation. Note that in
obtaining (\ref{ratio2}), we have used the freedom to reshuffle
subleading terms between numerator and denominator to cancel the sea
quark contributions common to $F_K$ and $F_\pi$, which are
proportional to the low energy constant (LEC) $L_4$. That leaves only
one NLO LEC in (\ref{ratio2}), i.e. $L_5(\mu)$, whose 
$\mu$ dependence cancels the one in the logarithms. Throughout,
we shall keep $\mu$ fixed to $770\MeV$, the value which is customary
in phenomenology.

In addition to \eq{ratio2}, we consider two more $SU(3)$ $\chi$PT
expressions, which are equivalent to (\ref{ratio2}) at NLO: the one
obtained by fully expanding (\ref{ratio2}) to NLO and the one obtained
by expanding the inverse of (\ref{ratio2}) to NLO. The use of these
three different forms is one of the ways that we have to estimate the
possible contributions of higher order terms (see \sec{sec:fitting}).

A number of collaborations have reported difficulties and large
corrections when fitting their results for $F_\pi$ and $F_K$ to NLO
$SU(3)$ chiral perturbation theory (or its partially quenched
descendent) around $M_\pi\gsim400\MeV$~\cite{Lellouch:2009fg}.  This
suggests that $m_s^\mr{phys}$ might be the upper end of the quark mass
range where NLO chiral perturbation theory applies. However, such a
statement may depend sensitively on the observable and in this respect
$F_K/F_\pi$ is rather special. It is an $SU(3)$-flavor breaking ratio
in which the sea quark contributions cancel at NLO. Moreover, the NLO
expressions fit our results for this ratio very well with only two
parameters, $F_0$ and $L_5$; the size of NLO corrections are perfectly
reasonable, of order 20\%; and the values of $F_0$ and $L_5$ that we
obtain  are acceptable for a fit performed at this 
order~\footnote{Note that the ratio $F_K/F_\pi$ is not well suited
  for a determination of $F_0$ and $L_5$. In this ratio, the LO LEC
  $F_0$ only appears at NLO where its value becomes highly correlated
  with that of $L_5$. A serious determination of these constants would
  actually require a dedicated study which is beyond the scope of this
  paper. }. In addition,
omission of all SU(3) fits would shift
our final result for $F_K/F_\pi$ by less than one quarter of our final
error, and slightly reduce the systematic error.

Ad ({\it ii}\,): Given the previous discussion,
it is clear that the heavy kaon $SU(2)$ framework of
\cite{Allton:2008pn} is an interesting alternative, since it does not
assume that $m_s/(4\pi F_0)\ll 1$ , but rather treats the strange
quark as a heavy degree of freedom, which is ``integrated out''.  Of
course, while this may be a good approximation at the physical value
of $m_{ud}$, where $m_{ud}/m_s\ll 1$, it may break down for larger
values, when $m_{ud}\lsim m_s$, as is the case for our more massive points.

What is needed for our analysis is Eq.\,(53) of \cite{Allton:2008pn}
for the pion mass dependence of $F_K$, together with the standard
$SU(2)$ formula for the mass dependence of $F_\pi$
\cite{Gasser:1983yg}. In these expressions, every low-energy constant
bears an implicit $m_s$ dependence which can be written as a power
series expansion in $m_s\!-\!m_s^\mr{phys}$ about the physical point. 
In practice we are only sensitive to the $m_s$ dependence of the 
LO term. Thus, we can express the ratio as
\beq
{\Fka\ovr\Fpi}=
{\bar F\ovr F}\;
\Big\{
1+\alpha\,\frac{m_s-m_s^\mr{phys}}{\mu_{\mathrm{QCD}}}
\Big\}\;
{1-{3\ovr8}{\Mpi^2\ovr(4\pi F)^2}\log\!\Big({\Mpi^2\ovr\bar\Lambda^2}\Big)\over
1-{\Mpi^2\ovr(4\pi F)^2}\log\!\Big({\Mpi^2\ovr\Lambda^2}\Big)}
\label{ratio3}
\ ,\eeq
where $F,\bar{F}$ denote the two-flavor chiral limit of the pion and
kaon decay constant, respectively, while $\Lambda,\bar\Lambda$ are the
energy scales associated with the respective LECs~\footnote{In the
  standard $SU(2)$ framework the former is usually denoted
  $\Lambda_4$, since it is associated with $l_4$
  \cite{Gasser:1983yg}.}. In (\ref{ratio3}), $\mu_\mr{QCD}$ is a
typical hadronic energy scale. 

This expression for $F_K/F_\pi$ has five $m_s$-independent parameters
(${\bar F}$, $F$, $\alpha/\mu_\mr{QCD}$, $\Lambda$, $\bar \Lambda$),
four if the ratio is expanded and only NLO terms are kept (${\bar F}$,
$F$, $\alpha/\mu_\mr{QCD}$, $\Lambda^{3/4}/\bar\Lambda^2$). We find
that the former leads to unstable fits and thus do not use it in our
analysis. One of the ways that we use to estimate possible contributions
of higher order terms (see \sec{sec:fitting}), is to consider two $SU(2)$
$\chi$PT expressions, which are equivalent to (\ref{ratio3}) at NLO:
the one, already discussed, obtained by fully expanding (\ref{ratio3})
to NLO and the other obtained by expanding the inverse of (\ref{ratio3})
to NLO.  For reasons similar to those discussed under ({\it i}\,), we
leave the determination of the LECs, which appear at NLO in
\eq{ratio3}, for future investigation.

In order to use the parameterization (\ref{ratio3}), one has to give a
definition of $m_s\!-\!m_s^\mr{phys}$. $SU(3)$ $\chi$PT, together with
\fig{fig1} and the discussion surrounding it, suggests that
$([2M_K^2-M_\pi^2]-[...]_\mr{phys})/\mu_\mr{QCD}\propto
(m_s\!-\!m_s^\mr{phys})[1+O(m_s\!-\!m_s^\mr{phys})]$, up to
negligible, physical, $m_{ud}$ and $m_{ud}^\mr{phys}$ corrections, in
the range of quark masses covered in our simulations. Thus, we use
this difference of meson masses squared as a measure of
$m_s\!-\!m_s^\mr{phys}$.

Ad ({\it iii}\,): Both two and three flavor chiral perturbation theory
are expansions about a singular point, $(m_{ud},m_s)=(0,0)$ in the
case of $SU(3)$ and $(m_{ud},m_s)=(0,m_s^\mr{phys})$ in the case of
$SU(2)$. However, here we are interested in the physical mass point,
$(m_{ud}^\mr{phys},m_s^\mr{phys})$, which is nearer the region in
which we have lattice results than it is to either singular
point. Thus, it makes sense to consider an expansion about a regular
point which encompasses both the lattice results and the physical
point~\cite{Lellouch:2009fg}. As discussed above, the data at two
lattice spacings ($\beta=3.3$ and $\beta=3.7$) have been generated at
a fixed value of $am_s$, and the data at $\beta=3.57$ can easily be
interpolated to a fixed $am_s$. On the other hand, an extrapolation is
needed in $m_{ud}$.  It is natural to consider the following
expansion parameters
\begin{eqnarray}
\Delta_\pi\! &=& \Big[
M_\pi^2 - \frac{1}{2}(M_\pi^\mr{phys})^2 - \frac{1}{2}(M_\pi^\mr{cut})^2
\Big]/\mu_\mr{QCD}^2
\\
\Delta_K\!   &=& \Big[
M_K^2-(M_K^\mr{phys})^2
\Big]/\mu_\mr{QCD}^2
\label{eq:deltaK}
\end{eqnarray}
with $\mu_\mr{QCD}$ as above and $M_\pi^\mr{cut}$ the heaviest
pion mass included in the fit. The discussion in
\cite{Lellouch:2009fg} suggests that the mass dependence of
$F_K/F_\pi$ in our ensembles and at the physical mass point can be
described by a low order polynomial in these variables, leading to the
``Taylor'' or ``flavor'' ansatz
\beq 
{\Fka\ovr\Fpi} = A_0 + A_1\Delta_\pi +
A_2\Delta_\pi^2 + B_1\Delta_K.
\label{ratio4}
\eeq
One of the ways that we use to estimate possible contributions of
higher order terms (see \sec{sec:fitting}), is to consider the same
functional form for $F_\pi/F_K$. Thus, we consider two flavor
ansaetze.

There are, of course, many possible variants of these ansaetze. For
instance, $\Delta_K$ in \eq{eq:deltaK} could be defined in terms of
$M_K$ instead of $M_K^2$. One could also enforce $SU(3)$ flavor
symmetry, i.e. $F_K/F_\pi=1$ when $m_{ud}=m_s$. This could be done,
for instance, with an expansion of the form
$F_K/F_\pi=1+(M_K^2-M_\pi^2)\times[ \mbox{polynomial in}$ $\Delta_\pi$
  and $\Delta_K]$. However, the expansion of \eq{ratio4} provides a
better description of our data than the alternatives that we have tried.


\subsection{Continuum limit}
\label{sec:contlim}

To maximize the use of our data, we combine the continuum
extrapolation with our interpolations and extrapolations to the
physical strange and up-down quark mass point.
As mentioned above, $F_K/F_\pi$ is an
$SU(3)$-flavor breaking ratio, so that cutoff effects must be
proportional to $m_s{-}m_{ud}$. Although \cite{Durr:2008rw}
suggests that they are quadratic in $a$, our action is
formally only improved up to $O(a\alpha_s)$ and we cannot exclude, a
priori, the presence of linear discretization errors. Moreover,
the effects are small enough that, despite a factor of almost two in
lattice spacing, we cannot distinguish $a$ from $a^2$ in our
data. Thus, in our analysis, we consider three possibilities: no
cutoff effects; cutoff effects of the form $a(m_s{-}m_{ud})$, and cutoff
effects proportional to $a^2\mu_\mr{QCD}(m_s{-}m_{ud})$.
Again, $\mu_\mr{QCD}$ is a typical hadronic energy scale.

In the case of $SU(3)$ $\chi$PT, the cutoff effects can be accounted for by
adding
\beq
\left.\frac{F_K}{F_\pi}\right\vert_\mr{c.o.} =c\times\left\{
\begin{array}{l}
a(M_K^2-M_\pi^2)/\mu_\mr{QCD}\\
\quad\mr{or}\\
a^2(M_K^2-M_\pi^2)\\
\end{array}
\right.
\label{eq:su3co}
\eeq
to the functional forms for the mass dependence of $F_K/F_\pi$ given
in the previous section. Here $c$ is the relevant parameter. 

In the usual power counting scheme of $SU(3)$ Wilson $\chi$PT, in
which $O(a)=O(m_q)$~\cite{Bar:2003mh}, the cutoff effects considered
in (\ref{eq:su3co}) are NNLO in the $O(a)$ case and even N$^3$LO for
those proportional to $a^2$. This should be compared to the physical
contributions to the deviation of $F_K/F_\pi$ from 1, which are
NLO. Thus, these cutoff effects are small parametrically and, as we
will see below, they are also small numerically. In fact, they are
consistent with zero within our statistical errors. Clearly then, any
reasonable parameterization of these small $SU(3)$-flavor breaking,
cutoff effects is sufficient for our purposes. Thus, we use the
parametrizations of \eq{eq:su3co} to subtract them, also in the
context of $SU(2)$ $\chi$PT and flavor expansions.


\subsection{Infinite volume limit}

The finite spatial size $L$ of the box affects masses and decay
constants of stable states in a manner proportional to $\exp(-\Mpi L)$
\cite{Luscher:1985dn}.  In our simulations the bound $\Mpi L\gsim 4$ is
maintained.
It turns out that the sign of the shift is the same for both decay
constants so that such effects partly cancel in the ratio. In chiral
perturbation theory to 1-loop order, $\Fpi(L)/\Fpi$ has been
calculated in \cite{Gasser:1986vb} and $\Fka(L)/\Fka$ in
\cite{Becirevic:2003wk}.  At the 2-loop level, both ratios have
been determined in \cite{Colangelo:2005gd}.  The generic
formulae take the form 
\bea {\Fka(L)\ovr\Fka}&=&1+\sum_{n=1}^\infty
{m(n)\ovr\sqrt{n}}\,{1\ovr\Mpi
L}\,{\Fpi\ovr\Fka}\,{\Mpi^2\ovr(4\pi\Fpi)^2}\,
\Big[I_{\Fka}^{(2)}+{\Mka^2\ovr(4\pi\Fpi)^2}I_{\Fka}^{(4)}+...\Big]
\\ {\Fpi(L)\ovr\Fpi}&=&1+\sum_{n=1}^\infty
{m(n)\ovr\sqrt{n}}\,{1\ovr\Mpi
L}\,\;\;1\;\;\,{\Mpi^2\ovr(4\pi\Fpi)^2}\,
\Big[I_{\Fpi}^{(2)}+{\Mpi^2\ovr(4\pi\Fpi)^2}I_{\Fpi}^{(4)}+...\Big]
\eea
with $m(n)$ tabulated in \cite{Colangelo:2005gd}.  With
$I_{\Fpi}^{(2)}=-4K_1(\sqrt{n}\,\Mpi L)$ and
$I_{\Fka}^{(2)}=-{3\ovr2}K_1(\sqrt{n}\,\Mpi L)$, where $K_1(.)$ is a
Bessel function of the second kind, one obtains the compact 1-loop
expression \cite{Colangelo:2005gd}
\beq
{\Fka(L)\ovr\Fpi(L)}={\Fka\ovr\Fpi}\; \bigg\{
1+\Big[4-{3\Fpi\ovr2\Fka}\Big]\sum_{n=1}^\infty
{m(n)\ovr\sqrt{n}}\,{1\ovr\Mpi L}\,{\Mpi^2\ovr(4\pi\Fpi)^2}\,
K_1(\sqrt{n}\Mpi L) \bigg\} \eeq
which is readily evaluated.  On the other hand, the expressions
$I_{\Fka}^{(4)}$, $I_{\Fpi}^{(4)}$ in the 2-loop part are more
cumbersome to deal with.  Since finite-volume effects are independent
of cutoff effects (see e.g.\ the discussion in
\cite{Colangelo:2005gd}), we choose to first correct the data for
finite volume effects before we actually perform the global fit (see
below). The masses of the mesons are also corrected for finite volume
effects according to~\cite{Colangelo:2005gd}. 

To estimate the error associated with the finite volume effects, the
difference between the 1-loop shift of $\Fka/\Fpi$ and the corrected
value of $\Fka/\Fpi$ using only the expression for $\Fpi$ (that should
be an upper bound on the finite volume correction), is used as a
measure of the uncertainty of the correction (see \sec{sec:fitting}).
In our final analysis, the finite volume shift is smaller than the
statistical error in most of our ensembles. 


\subsection{Fitting strategy and treatment of theoretical errors}
\label{sec:fitting}

Our goal is to obtain $F_K/F_\pi$ at the physical point, in the
continuum and in infinite volume.  To this end we perform a global fit
which simultaneously extrapolates or interpolates
$\Mpi^2\to\Mpi^2|_\mr{phys}$, $\Mka^2\to\Mka^2|_\mr{phys}$ and
$a\to0$, after the data have been corrected for very small finite
volume effects, using the two-loop chiral perturbation theory results
discussed above. To assess the various systematic uncertainties
associated with our analysis, we consider a large number of alternative
procedures.

To estimate the systematic uncertainty associated with setting the
scale, we have repeated the analysis setting the scale with the $\Xi$
and the $\Omega$.

To estimate the possible contributions of excited states to the
correlators used, we repeat our analysis using a total of $18$
different time intervals: $t_\mr{min}/a=5$ or 6, for $\beta=3.3$; 7, 8
or 9, for $\beta=3.57$; 10, 11 or 12 for $\beta=3.7$. All of these
intervals are chosen so that the correlation functions are strongly
dominated by the ground state. However, the intervals which begin at
earlier times may receive some small contamination from excited
states.

Uncertainties associated with the necessary extrapolation to the
physical up and down quark-mass point are assessed by varying the
functional form used as well as by varying the range of pion masses
considered. As discussed in \sec{sec:massextrap}, we consider 3 forms
based on the NLO $SU(3)$ formula of \eq{ratio2}, 2 forms based on the
NLO $SU(2)$ expression of \eq{ratio3}, and 2 flavor ansaetze,
based on \eq{ratio4}. Moreover, we impose 2 cuts on the pion mass:
$M_\pi\!<\!350\MeV$ and $460\MeV$. 

Our results for $F_K/F_\pi$ display a small dependence on lattice
spacing. As described in \sec{sec:contlim}, to estimate the systematic
associated with the continuum extrapolation we consider fits with and
without $O(a^2)$ and $O(a)$ Symanzik factors.

All of this amounts to performing $2\cdot 18\cdot 7\cdot 2\cdot
3=1512$ alternative analyses.  The central value obtained from each
procedure is weighted with the quality of the (correlated) fit to
construct a distribution. The median and the 16-th/84-th percentiles
yield the final central value and the systematic error associated with
possible excited state contributions, scale setting, and the chiral
and continuum extrapolations.  To obtain the final systematic error of
our computation, we add a finite-volume uncertainty in quadrature to
the error already obtained. This finite-volume uncertainty is given by
the weighted (with the quality of the fit) standard deviation of our
final central value and the ones obtained repeating the whole
procedure with finite-volume effects subtracted at one loop and the
upper bound computed as the 2 loop correction in $F_\pi$ only. We do
not include these two alternative options in the set of procedures
which yield our final central value because we know, a priori, that
the two-loop expressions of \cite{Colangelo:2005gd} are more accurate
than the one-loop ones. To determine the statistical error, the whole
procedure is bootstrapped (with 2000 samples) and the variance of the
resulting medians is computed.

There is a final source of theoretical error which we wish to comment
on: the one associated with the fact that our calculation is performed
with $m_u{=}m_d$ and in the absence of electromagnetism. As discussed
in the Introduction, we correct for these effects at leading order, up
to a few per mil uncertainties. The latter have a negligible effect on
$F_K/F_\pi$. One also expects direct isospin violation in $F_K/F_\pi$.
These were found to be negligible in \cite{Aubin:2004fs}.

Having estimated the total systematic error, it is interesting to
decompose it into its individual contributions. By construction, the
contribution from the uncertainty in the finite-volume corrections
is already known. To quantify the other contributions, we
construct a distribution for each of the possible alternative
procedures corresponding to the source of theoretical error under
investigation. These distributions are obtained by varying over all of
the other procedures and weighing the results by the total fit
quality. Then, we compute the medians and average fit qualities of these
distributions. Clearly the spread of the medians measure the uncertainty
associated with the specific source of error. We use the standard
deviation of these medians weighted by the average fit quality as
an estimate of the error under consideration.

\begin{figure}[t,b]
\epsfig{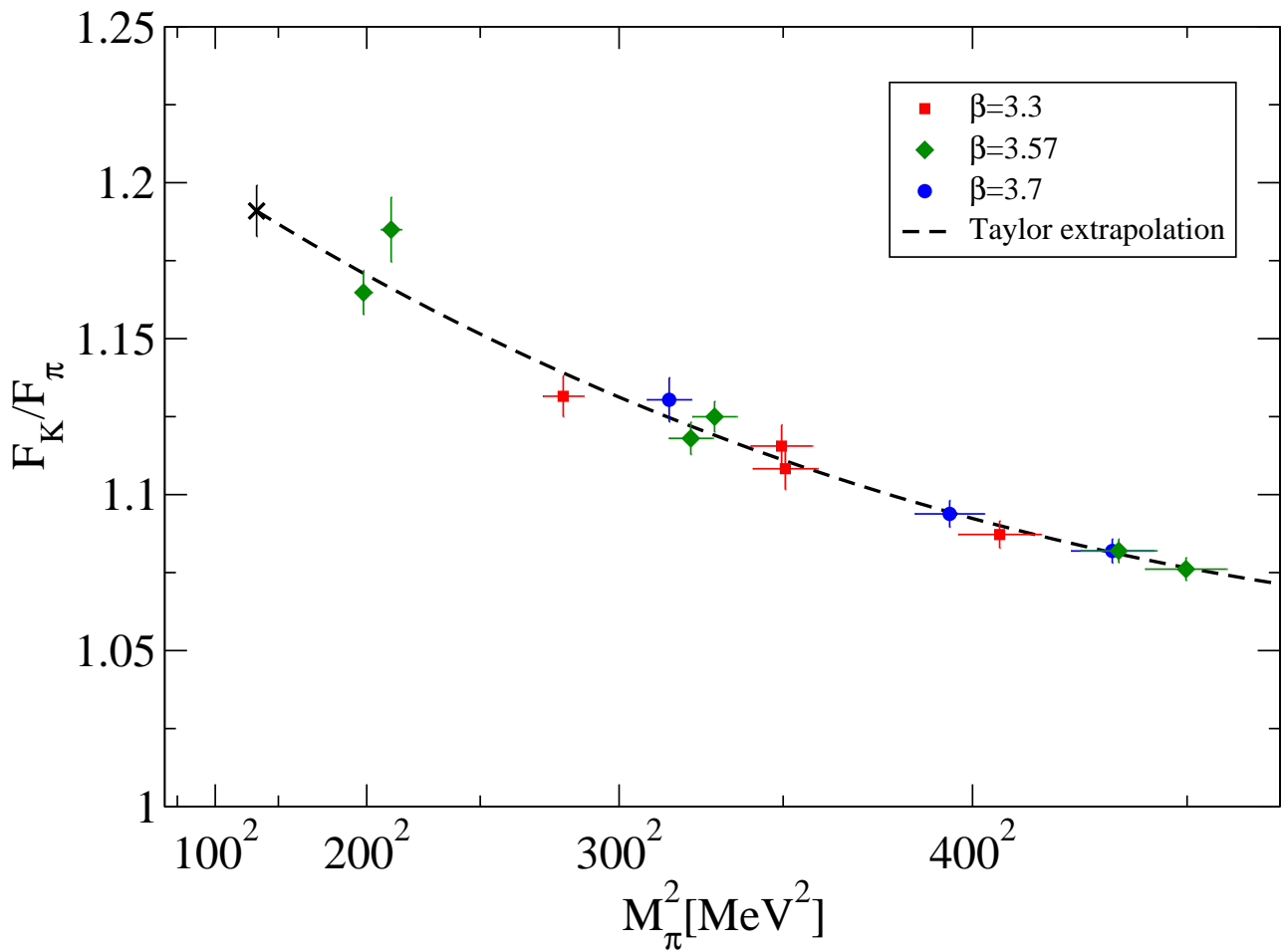}
\vspace{-8mm}
\caption{\sl Extrapolation of the lattice data to the physical
  point for a particular choice of two-point function fits
  ($t_{\mr{min}}/a= 6,8,11$ for $\beta=3.3,3.57,3.7$, respectively),
  mass cut ($M_\pi < 460 \MeV$) and using the $\Xi$ to set the scale.
  The plot shows one (of the 21) fits used 
  to estimate the uncertainty associated with the functional form used
  for the mass extrapolation. The data have been
  slightly adjusted to the physical
  strange quark mass, as well as corrected for tiny finite-volume
  effects (see text for details). }
\label{fig2}
\end{figure}

A ``snapshot'' fit (with a
specific choice for the time intervals used in fitting the
correlators, scale setting, pion mass range) can be seen
in~\fig{fig2}. To avoid the complications of a multi-dimensional 
plot, the extrapolation is shown as a function of the pion mass only.
The data have been corrected for the deviation of the simulated $m_s$
from $m_s^\mr{phys}$.  In other words, what is shown is
$\mr{data}(\Mpi^2,2\Mka^2\!-\!\Mpi^2)-\mr{fit}(\Mpi^2,2\Mka^2\!-\!\Mpi^2)+
\mr{fit}(\Mpi^2,[2\Mka^2\!-\!\Mpi^2]_\mr{phys})$. The
figure shows one flavor fit with no cutoff effects. We emphasize that 
$\chi^2/dof$ for our correlated fits are close to one.

\begin{figure}[t,b]
\epsfig{file=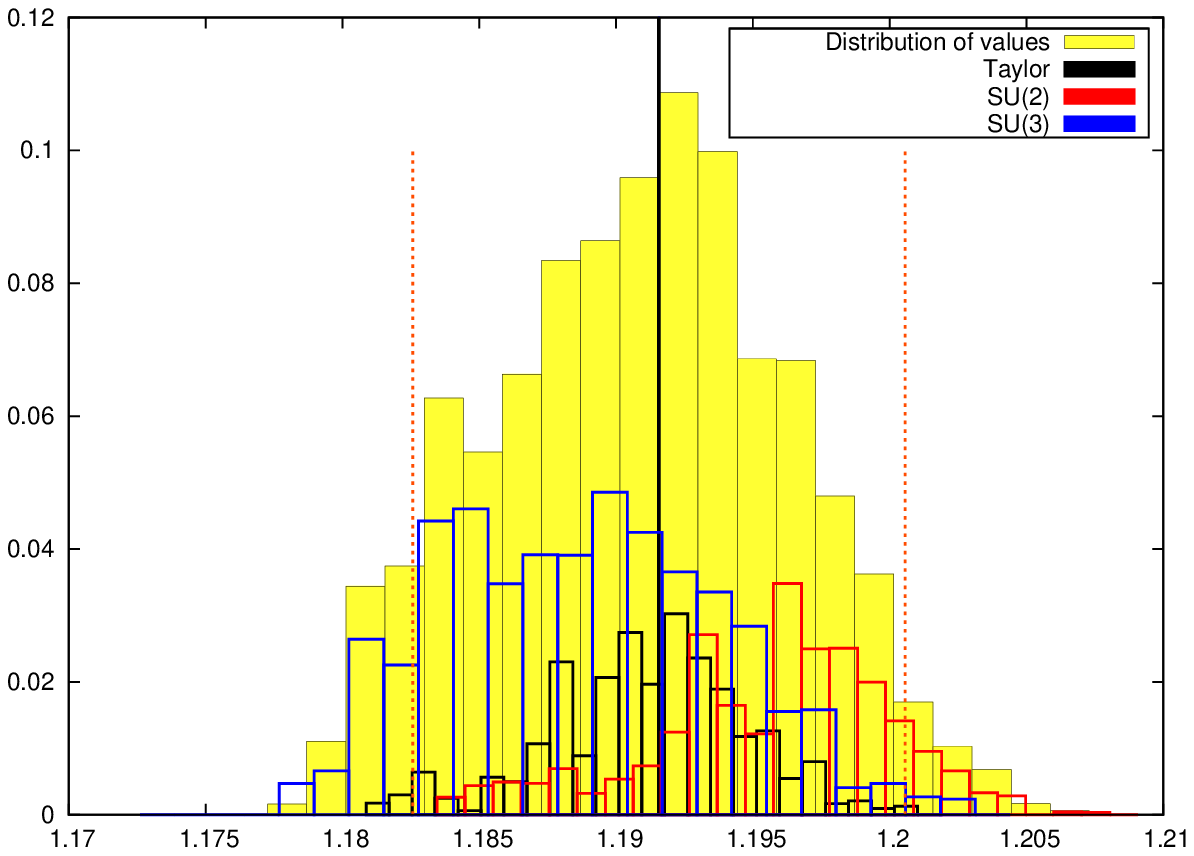,width=17cm}
\vspace{-8mm}
\caption{\sl Distribution of final values for $F_K/F_\pi$. The
  large background distribution represents the values of $F_K/F_\pi$
  obtained with different extrapolation formulas, pion mass cuts,
  parameterization of cutoff effects, time intervals and different
  methods to set the scale. Also shown is the final result (solid
  vertical line) and final systematic error interval (dashed vertical
  lines), which includes the finite volume error estimate.
  The smaller distributions group the values
  according to the chiral formula used for the extrapolation. The
  error associated with the chiral extrapolation is computed as the
  weighted (by total fit quality) standard deviations of the medians
  of the grouped distributions.}
\label{figmth}
\end{figure}

\fig{figmth} shows the distribution of our 1512 alternative fitting
procedures. As can been seen, the use of different formulas
to extrapolate to the physical point only increases the systematic
error. This is also true for the other sources of systematic error:
scale setting, time intervals for the fit to the correlators, pion
mass ranges and cutoff effects. The figure also shows our final result
with the total final error.


\section{Results and discussion}


Following the procedure outlined above, our final result is
\beq
{F_K\over F_\pi}\bigg|_\mr{phys}=1.192(7)_\mr{stat}(6)_\mr{syst}
\qquad\mbox{or}\qquad
{F_\pi\over F_K}\bigg|_\mr{phys}=0.839(5)_\mr{stat}(4)_\mr{syst}
\label{final_ratio}
\eeq
at the physical point, where all sources of systematic error have
been included. \tab{tab:err} gives a breakdown of the systematic error
according to the various sources. We conclude that our main source of
systematic error comes from the chiral extrapolation (functional form
and pion mass range), followed by cut-off effects.
\begin{table}
  \centering
  \begin{tabular}{lc}
    \hline
    \hline
    Source of systematic error & {error on $F_K/F_\pi$} \\
    \hline
    Chiral Extrapolation: &  \\
     - Functional form & $3.3\times 10^{-3}$ \\
     - Pion mass range & $3.0\times 10^{-3}$ \\
    Continuum extrapolation & $3.3\times 10^{-3}$ \\
    Excited states & $1.9\times 10^{-3}$ \\
    Scale setting & $1.0\times 10^{-3}$\\
    Finite volume & $6.2\times 10^{-4}$\\
    \hline
    \hline
  \end{tabular}
  \caption{\sl Breakdown of the total systematic error on $F_K/F_\pi$
    into its various components, in order of decreasing
    importance. }
  \label{tab:err}
\end{table}

In the same manner we may read off from our fits that the value in the
2-flavor chiral limit \beq {F_K\over
  F_\pi}\bigg|_{m_{ud}=0,m_s\,\mr{phys}}=1.217(10)_\mr{stat}(9)_\mr{syst}
\qquad\mbox{or}\qquad {F_\pi\over
  F_K}\bigg|_{m_{ud}=0,m_s\,\mr{phys}}=0.821(7)_\mr{stat}(6)_\mr{syst}
\eeq differs by about 2\% from the physical value, though with larger
uncertainties.  Upon combining this ratio in the $SU(2)$ chiral limit
with the phenomenological value $F=\lim_{m_{ud}\to0}F_\pi=86.2(5)\MeV$
\cite{Colangelo:2003hf}, we obtain
$\lim_{m_{ud}\to0}F_K=104.9(1.3)\MeV$ at fixed physical $m_s$.

\begin{figure}[t]
\centering
\epsfig{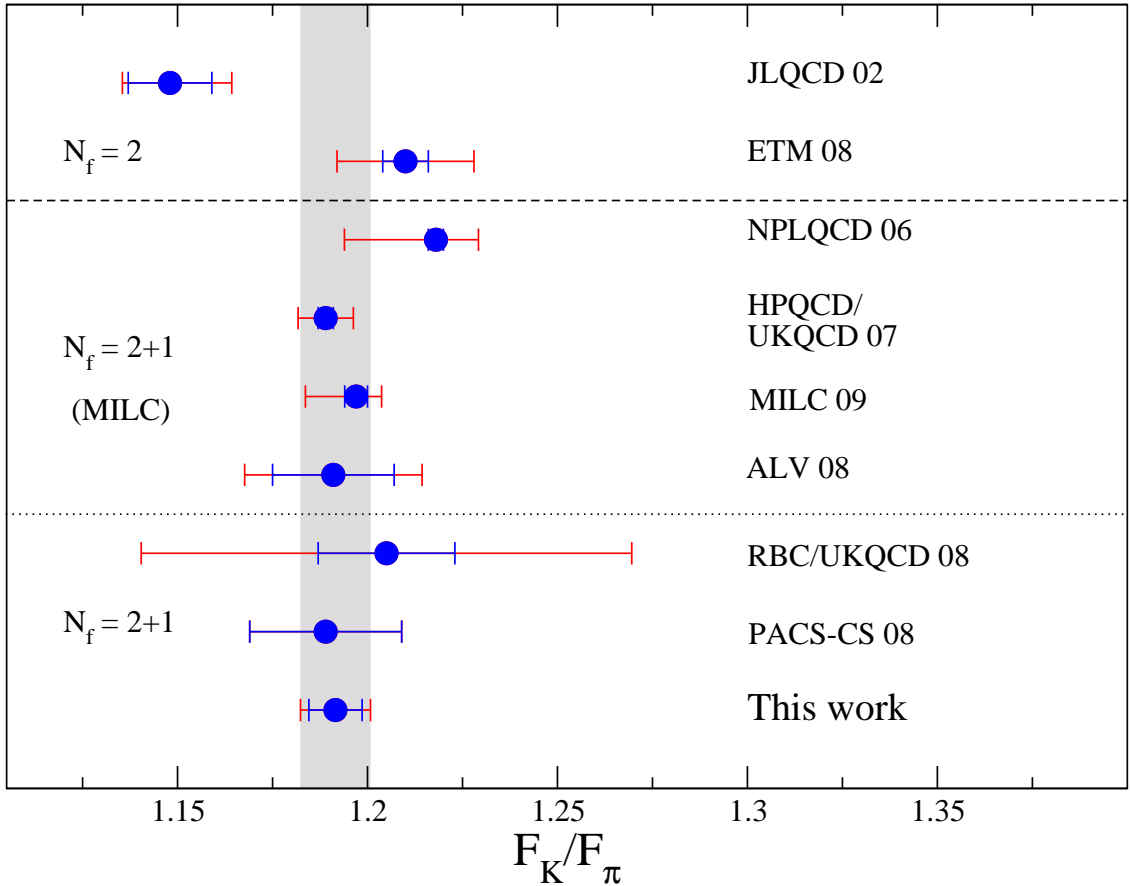}
\caption{\sl Our result (\ref{final_ratio}) compared to previous unquenched
lattice QCD calculations. The two error bars refer to the statistical and
to the combined statistical and systematic uncertainties,
respectively. The top two results were obtained with $N_f=2$
simulations. The next four were computed on $N_f=2+1$ MILC
configurations. RBC/UKQCD and PACS-CS determinations were obtained
on distinct $N_f=2+1$ ensembles.}
\label{fig4}
\end{figure}

In \fig{fig4} a survey of recent determinations of $F_K/F_\pi$ in
unquenched QCD simulations is presented.
These include results by
JLQCD \cite{Aoki:2002uc},
MILC \cite{Aubin:2004fs,
           Bazavov:2009bb},
NPLQCD \cite{Beane:2006kx},
HPQCD/UKQCD \cite{Follana:2007uv},
ETM \cite{Blossier:2009bx}
RBC/UKQCD \cite{Allton:2008pn}, PACS-CS \cite{Aoki:2008sm}, and Aubin
et al.\ \cite{Aubin:2008ie}.  It is worth noting that these results
show a good overall consistency when one excludes the outlier point of
\cite{Aoki:2002uc}.

The result by HPQCD/UKQCD remains the one with the highest claimed
accuracy.  Our precision is similar to that of another
state-of-the-art calculation, the one by MILC. Given that we reach
smaller pion masses, that more generally we explore a large range of
simulation parameters which allows us to control all sources of
systematic error, that we use an action with a one-to-one matching of
lattice-to-continuum flavor and that we avoid the specifics of a
partially quenched framework, our result solidifies the claim that
$F_K/F_\pi$ is known to better than 1\%. Moreover, agreement between
such different approaches can only bolster confidence in
the reliability of lattice calculations per se.


\section{Update on $|V_\mr{us}|$ and check of CKM unitarity relation}


With the result (\ref{final_ratio}) in hand, we can now focus on CKM matrix
elements.
In this respect there are two options -- we may \emph{assume} the SM (and hence
CKM unitarity) and determine $|V_\mr{ud}|$ and $|V_\mr{us}|$, or we may use
phenomenological input on $|V_\mr{ud}|$ to derive $|V_\mr{us}|$ and hence
\emph{test} CKM unitarity (under the assumption of quark-flavor universality)
in a model-independent way.

The first step, needed in either case, is to simplify Marciano's
\eq{marciano}.   
The most recent update of the Flavianet kaon working group is
\cite{Antonelli:2008jg}
\beq
{|V_\mr{us}|\ovr |V_\mr{ud}|}\,{F_K\ovr F_\pi}=0.27599(59)
\;.
\eeq
Combining this with our result (\ref{final_ratio}) yields the ratio
\beq
|V_\mr{us}|/|V_\mr{ud}|=0.2315(19)
\;.
\label{final_Vus/Vud}
\eeq
Now for the two options.
If we assume unitarity, (\ref{final_Vus/Vud}) and
$|V_\mr{ub}|\!=\!(3.93\!\pm\!0.36)\,10^{-3}$ \cite{Amsler:2008zz} imply
\beq
|V_\mr{ud}|=0.97422(40)
\;,\quad
|V_\mr{us}|=0.2256(17)
\;.
\eeq
On the other hand, if we combine (\ref{final_Vus/Vud}) with the most precise
information on the first CKM matrix element available today,
$|V_\mr{ud}|=0.97425(22)$ \cite{Hardy:2008gy}, we obtain (again)
\beq
|V_\mr{us}|=0.2256(18)
\;.
\label{final_Vus}
\eeq
Similarly, by also including the above mentioned result for $|V_\mr{ub}|$ we
find
\beq
|V_\mr{ud}|^2+|V_\mr{us}|^2+|V_\mr{ub}|^2=1.0001(9)
\;.
\label{final_sum}
\eeq
With the first-row unitarity relation (which is genuine to the CKM paradigm)
being so well observed, there is no support for ``beyond the Standard Model''
physics contributions to these processes.
Of course, with substantially improved precision on both the theoretical and
the experimental side, this might change in the future.


\section{Summary}


Based on a series of 18 large scale lattice computations the ratio
$F_K/F_\pi$ has been determined in $N_f\!=\!2\!+\!1$ flavor QCD at the
physical mass point, in the continuum and in infinite volume.  The
overall precision attained is at the 1\% level, with statistical and
systematic errors being of similar magnitude.  The systematic error,
in turn, splits into uncertainties arising from the extrapolation to
the physical pion mass, to the continuum, to possible excited state
contributions and to finite volume effects. We find that the main
source of systematic error comes from the extrapolation to the
physical point.  In this respect it is useful that our simulations
cover the pion mass range down to about $190\MeV$, with small cutoff
effects and volumes large enough to maintain the bound $\Mpi L \gsim
4$.

Following Marciano's suggestion \cite{Marciano:2004uf}, we use our
result (\ref{final_ratio}) for $F_K/F_\pi$ to obtain the value
(\ref{final_Vus/Vud}) for $|V_\mr{us}|/|V_\mr{ud}|$ and subsequently
$|V_\mr{us}|$. In turn, these results are used to test first row
unitarity, which we find is satisfied at the 1 per mil level, thereby
imposing interesting contraints on ``new physics'' scenarios (see
e.g. \cite{Antonelli:2008jg}).


\bigskip
\noindent
{\bf Acknowledgments}:
Computations were performed on Blue Gene supercomputers at FZ J\"ulich and
IDRIS (GENCI), as well as on clusters at Wuppertal and CPT.
This work is supported in part by the U.S.\ Department of Energy under Grant
No.\ DE-FG02-05ER25681, EU grant I3HP, OTKA grant AT049652, DFG grant
FO 502/1-2, SFB/TR-55, EU RTN contract MRTN-CT-2006-035482 (FLAVIAnet) and by
the CNRS's GDR grant n$^o$ 2921 and PICS grant n$^o$ 4707.

\clearpage



\end{document}